\documentclass[twocolumn,showpacs,preprintnumbers,amsmath,amssymb]{revtex4}
\usepackage[dvips]{graphicx}
\usepackage{dcolumn}
\usepackage{bm}

\def\ket#1{ \left\vert  #1  \right\rangle }
\def\bra#1{ \left\langle  #1 \right\vert}
\def\spr#1#2{ \left\langle #1 \left\vert \right. #2 \right\rangle }
\def\etal{\textit{et al.}}

\begin{document}

\title{Quantum computing with mixed states}
\author{Michael Siomau}
\email{siomau@physi.uni-heidelberg.de}
\affiliation{Max-Planck-Institut f\"{u}r Kernphysik, Postfach
             103980, D-69029 Heidelberg, Germany}
\affiliation{Physikalisches Institut, Heidelberg Universit\"{a}t,
D-69120 Heidelberg, Germany}

\author{Stephan Fritzsche}
\affiliation{Department of Physical Sciences, P.O.~Box 3000,
             Fin-90014 University of Oulu, Finland}
\affiliation{GSI Helmholtzzentrum f\"{u}r Schwerionenforschung,
D-64291 Darmstadt, Germany}

\begin{abstract}
We discuss a model for quantum computing with initially mixed
states. Although such a computer is known to be less powerful than a
quantum computer operating with pure (entangled) states, it may
efficiently solve some problems for which no efficient classical
algorithms are known. We suggest a new implementation of quantum
computation with initially mixed states in which an algorithm
realization is achieved by means of optimal basis independent
transformations of qubits.
\end{abstract}


\maketitle

\section{Introduction}

Quantum computers hold the promise to solve feasible computational
problems requiring excessive resources for their solution on
classical computers. Despite the large number of profound results
\cite{Jozsa:03,Blume:02,Linden:01,Braunstein:02,Biham:04}, the
question: ``What is the essential quantum effect that gives rise to
increase in the computing power of a quantum computer?'' -- remains
the most enigmatic in the theory of quantum computation. A partial
answer has already been given by Jozsa and Linden \cite{Jozsa:03}
who proved that for any quantum algorithm operating with {\it pure}
states the presence of multi-partite {\it entanglement} is necessary
to offer a computational speed-up over classical computing. However,
it has also been pointed out that a computational speed-up is
possible for algorithms operating with {\it mixed} states even in
the total absence of entanglement. Hitherto, there is no common
agreement on the role of entanglement and `mixedness' of quantum
states in the quantum computation theory \cite{Horodecki:09}.

In general, a quantum computer is a device that runs a program
through a carefully controlled sequence of unitary operations
(and/or measurements) applied to initially prepared states of
quantum systems. The answer is stored as classical information that
can be read out with high probability by a measurement. To be more
specific in the definition of a quantum computer, DiVincenzo
formulated the five requirements for the architecture and physical
implementation of a quantum computer \cite{DiVincenzo:00}, such as:
\begin{itemize}
 \item {\it Scalability.} A scalable physical system with well
characterized parts, usually qubits -- two-level quantum systems, is
available.
 \item {\it Initialization.} It is possible to prepare the
system in a simple state, such as $\ket{00...0}$.
 \item {\it Control.} Control of a quantum computation is
accomplished via some universal set of elementary unitary
operations.
 \item {\it Stability.} The system has long relevant
decoherence time, much longer than the gates times.
 \item {\it Measurement.} It is possible to read out the state of
the computer in a convenient product basis called the computational
basis.
\end{itemize}
If a quantum computer satisfies the five requirements above, it is
called a scalable quantum computer (SQC). For such a computer the
multi-qubit entanglement of pure states is proven to be the source
of computational power \cite{Jozsa:03,Blume:02}.

However, first experiments on the realization of a SQC
\cite{Cirac:95,Pellizzari:95,Gershenfeld:97} faced many difficulties
mostly connected to the control and the stability of the quantum
systems. Indeed, real quantum systems are rarely in pure states and
continuously interact with their environments which lead to
non-unitary (uncontrolled and unstable) evolution. Furthermore, the
proposals and experiments using NMR at high temperature to study
quantum computation \cite{Gershenfeld:97} involve manipulations with
initially mixed states giving rise to the problem of the
initialization of the system. Since fully controllable and scalable
quantum computers are still quite a way in the future, a less
ambitious quantum processor, that may fail to satisfy one or more of
the five criteria above but can nonetheless carry out interesting
computations, is of great interest. The first investigation of the
power of a quantum computer that breaks the second (the
initialization) requirement was presented by Knill and Laflamme
\cite{Knill:98}, who discussed deterministic quantum computation
with just one qubit in an initially mixed state. The computation
with the mixed state was shown to be less powerful than a SQC.
However, some problems related to physical simulations, for which no
efficient classical algorithms are known, can be solved with its
help. In general, the model of quantum computing with mixed states
is recognized to be somewhere in between classical computers and
SQCs \cite{Blume:02}.

Any model for quantum computation can be viewed to include three
parts, such as {\it quantum algorithms}, {\it an architecture} and
{\it physical realizations of a quantum computer}. In this work we
shall discuss step by step the three parts of the model for quantum
computation with initially mixed states, focusing especially on the
possible architecture of such a computer. To keep the discussion
precise, we shall restrict ourselves with the deterministic circuit
model for quantum computation \cite{Nielsen:00}. In this model any
algorithm can be represented through a sequence of (deterministic)
logic gates acting on initially prepared states, and just a finite
set of gates is sufficient to perform an arbitrary quantum algorithm
\cite{DiVincenzo:95}. We shall establish a universal set of quantum
gates that may efficiently operate with mixed states, in the sense
that the loss of information during the computation is less then a
given value $\delta$. These gates provide optimal basis independent
transformations of input qubits and are shown to be strongly related
to the optimal {\it cloning transformations} \cite{Scarani:05}.

This work is organized as follows. In the next section we shall very
briefly analyze some of the existing quantum algorithms
\cite{Shor:94,Grover:97,Deutsch:92,Simon:97} from the viewpoint of
their computational complexity in order to figure out possible
advantages of quantum computing with mixed states over classical
computation. It is shown that, apart from the quantum computation
with a single qubit in a mixed state \cite{Knill:98}, the
Deutsch-Jozsa \cite{Deutsch:92} and the Simon \cite{Simon:97}
problems can be solved more reliably by means of quantum computing
with mixed states then by the best possible classical algorithm. We
start Section~\ref{sec:3} with the analysis of the common
architecture for quantum computation \cite{Nielsen:00} in which a
universal set of usual basis dependent gates is used in order to
provide an arbitrary computation. While this scheme is shown to be
hardly suitable to support a quantum computation with initially
mixed states, we suggest and analyze in Sections~\ref{sec:3.2} and
\ref{sec:3.3} a new (so-called `cloning-based') architecture of
quantum computation. Section~\ref{sec:4} is devoted to a brief
analysis of possible physical realizations of a SQC as well as the
suggested cloning-based scheme for a quantum computer operating with
mixed states.

\section{\label{sec:2} On computational complexity of quantum algorithms}

Before starting any discussion about the architecture and physical
realization of a quantum computer operating with mixed states, it is
necessary to understand for which computational problems such a
computer may provide advantages compared to a classical one. Without
trying to analyze all existing quantum algorithms, we consider here
just some of them. The most exciting quantum algorithm, that
provides an {\it exponential speed-up} over the best {\it known}
classical algorithm for factoring problem, was suggested by Shor
\cite{Shor:94}. According to Shor's original construction, this
algorithm starts with pure (separable) states and requires highly
entangled pure states during its realization \cite{Shor:94}.
Interestingly, although the multi-qubit entanglement of pure states
was shown to be necessary to achieve exponential speed-up over
classical computing \cite{Linden:01}, it is still not clear what is
the sufficient condition to reach this computational advantage
\cite{Horodecki:09}. In practice, moreover, it is impossible to
avoid decoherence of the pure entangled states that are involved in
a computation. It was shown by Palma with co-authors \cite{Palma:97}
that decoherence decreases the probability of successful computation
of Shor's algorithm exponentially with the length of input data.
This result implies that the computation of Shor's algorithm with
initially mixed states has an exponentially small advantage over a
classical computation.

Another practically important quantum algorithm was suggested by
Grover \cite{Grover:97}. With the help of this algorithm, a search
through an unstructured database can be performed with a {\it
quadratic speed-up} over the best \textit{possible} classical
algorithm. For Grover's algorithm the multi-qubit entanglement of
pure states was shown \cite{Braunstein:02} to be necessary to obtain
a computational advantage over classical computing. It was also
shown that the computational speed-up using mixed states is not
possible except for the special case of the search space of size
four.

Recently, Biham \etal{} \cite{Biham:04} analyzed how fast the
Deutsch-Jozsa \cite{Deutsch:92} and the Simon \cite{Simon:97}
problems can be implemented with a quantum computer operating with
initially mixed states. It was found that these quantum algorithms
can be solved more reliably by means of quantum computing with mixed
states then by the best \textit{possible} classical algorithm. For
an arbitrary pure n-qubit state $\ket{\psi}$ and real (purity)
parameter $0 \leq \epsilon \leq 1$, it was proven that quantum
computation with the \textit{pseudo-pure} state $ \rho = \epsilon
\ket{\psi} \bra{\psi} + (1 - \epsilon) I^{\otimes n} $, where $I$
denotes the identity operator of the second order, guarantees a
speed-up over classical algorithms even when the purity parameter
$\epsilon$ is arbitrarily close to zero. However, the speed-up of
the quantum algorithms rapidly decrease with the number of qubits
involved in the computation \cite{Biham:04}.

Finally, a quantum computation with a single qubit pseudo-pure state
has advantages over classical computing \cite{Knill:98}, as
mentioned in the introduction. For a more detailed discussion about
computing with a single qubit mixed state we refer to the recent
contribution by Shepherd \cite{Shepherd:10}.

Although the question of the existence of a non-vanishing advantage
of quantum computing with mixed states is still open
\cite{Jozsa:03,Biham:04}, the brief analysis above shows that this
type of quantum computing may support classically unavailable
information processing. In the next section, therefore, we analyze
the possible architecture for effective quantum computing with mixed
states. As a particular form of the mixed states, moreover, we shall
use pseudo-pure states as they are given above.

\section{\label{sec:3} Architecture of quantum computer}

Representing internal operational structure of a quantum computer,
an architecture necessarily includes a finite {\it universal} set of
elementary operations that need to be performed on initially
prepared states of quantum systems according to rigourously defined
rules. In the quantum circuit model, the universal set of operations
may consist of single qubit gates with a two-qubit {\it
Controlled-NOT} gate or, alternatively, of single qubit gates with a
multi-qubit {\it Toffoli} gate \cite{Nielsen:00}. Universality of
the set implies that an arbitrary quantum algorithm can be
implemented with the help of the operations from the set.

\subsection{\label{sec:3.1} Scalable quantum computer}

A well-known architecture for a SQC within the circuit model for
quantum computing includes a universal set of gates consisting of
single qubit gates with a two-qubit C-NOT gate. These gates are
usually assumed to be implemented on initially pure states of qubits
and are associated within a specific computational basis. The C-NOT
gate requires two input qubits, one of which is called the control
and the other -- the target; it is usually defined in the
computational basis as \cite{Nielsen:00}
\begin{equation}
\label{standart-CNOT}
 U = \ket{0}\bra{0}_c\otimes I_t + \ket{1}\bra{1}_c \otimes
 (\sigma_x)_t \, ,
\end{equation}
where $\sigma_x = \ket{0}\bra{1} +  \ket{1}\bra{0}$. This gate
(\ref{standart-CNOT}) creates entanglement between initially
separable input (pure) states of control and target qubits, if the
control qubit is given in a superposed state of the basis states
$\ket{0}_c$ and $\ket{1}_c$. For example, if the input control and
target qubits are given in the states $\ket{+}_c = \sqrt{1/2}
(\ket{0} + \ket{1})$ and $\ket{0}_t$ respectively, the output state
is the maximally entangled (Bell) state $\ket{\phi} = \sqrt{1/2}
(\ket{00} + \ket{11})$ of two qubits.

Suppose, the input states of the control and the target qubits are
not pure anymore but pseudo-pure. Assume, for simplicity, that the
input state of the control qubit is given by $ \rho_c = x \ket{+}
\bra{+} + (1 - x ) I$ while the input state of the target qubit is $
\rho_t = x \ket{0} \bra{0} + (1 - x ) I$ with an equal purity
parameter $x$. Applying the C-NOT gate (\ref{standart-CNOT}) to the
input qubits in the mixed states, let us analyze how rapid the
entanglement of the output two-qubit state decreases with regard to
the purity $x$ of the input states. To quantify entanglement of the
output two-qubit state we use an entanglement measure -- {\it
concurrence} as suggested by Wootters \cite{Wootters:98}. We found
that the concurrence for the output two-qubit state decreases with
the purity parameter as $C = {\rm max} \{ 0, \, 1/2 (x^2 + 2 x - 1)
\}$. While it is often required to apply the C-NOT gate
(\ref{standart-CNOT}) many times during a computation, the
significant loss of entanglement of the output state after a single
C-NOT operation makes impossible an effective quantum computation
with input mixed states within the scheme for quantum computation
with basis dependent gates. Moreover, the computation with
pseudo-pure states is not possible at all for the input states with
the purity $x < 0.414$, since the concurrence for the output
two-qubit state from the C-NOT gate vanishes.

\subsection{\label{sec:3.2} Cloning-based architecture:
                            single-qubit gates}

Since the early days of the development of the theory for quantum
computation, quantum logical gates are associated with the
computational basis. Nowadays, this assumption is widely accepted,
but is not indeed necessary. It is possible to define quantum gates
to be {\it basis independent}. Several examples of optimal basis
independent operations have already been discussed in the
literature, such as universal NOT \cite{Buzek:99,Gisin:99},
universal Hadamard \cite{Pati:02} and universal C-NOT
\cite{Siomau:10} gates. Here we note that the optimal basis
independent operations are usually referred to universal quantum
operations in the literature \cite{Scarani:05}, in the sense that
they provide specific transformations on input qubits independently
from their initial states and with maximal possible fidelity between
the output states and the `idealized' output states. With the help
of mentioned optimal basis independent gates, a universal set of
gates, which is sufficient to provide an arbitrary quantum
computation, can be constructed. In the next two sections,
therefore, we present our central result -- the universal set of
basis independent gates consisting of single qubit gates and a
multi-qubit Toffoli gate. Consequently, we show that the optimal
basis independent gates may efficiently operate with initially mixed
states.

Let us get started with single qubit universal gates. The most
important single qubit universal gate is the NOT gate that generates
at the output the orthogonal state $\ket{\psi^\bot}$ with regard to
the input state $\ket{\psi}$, where $\spr{\psi} {\psi^\bot} \equiv
0$ for an arbitrary input qubit state $\ket{\psi}$. It was shown
that the exact universal NOT gate for an arbitrary input qubit state
does not exist, while it is possible to provide the corresponding
transformation approximately \cite{Buzek:99}. At the same time, it
is possible to construct an exact universal NOT gate for input
states taken from a restricted (but still infinite) set of states
\cite{Pati:00}. For example, input states may be chosen from a
one-dimensional subspace of the two-dimensional Hilbert space of
qubit states. Using the (Poincare-)Bloch sphere representation of a
qubit, a one-dimensional subspace can be visualized as an
intersection of the sphere with a plane. Let us consider the
one-dimensional subspace, \textit{the main circle}, that is formed
by the intersection of the Bloch sphere with the \textit{x-z} plane.
An arbitrary qubit state in this circle can be parameterized as
\begin{eqnarray}
 \label{states}
\ket{\psi} & = & \cos{\frac{\theta}{2}} \ket{0} \pm
\sin{\frac{\theta}{2}}\ket{1} \,
\end{eqnarray}
and is called a {\it real} qubit state. The meridional angle
$\theta$ takes values $0 \leq \theta \leq \pi$. For an arbitrary
input state (\ref{states}) the gate
\begin{equation}
\label{NOT}
    {\rm NOT} = - i \sigma_y = \left(
\begin{array}{cc} 0 & -1 \\ 1 & 0 \end{array} \right) \,
\end{equation}
provides the exact basis independent NOT operation, i.e. $\spr{\psi}
{{\rm NOT} |\, \psi} \equiv 0$ \cite{Pati:00,Siomau:10}.

With the help of the NOT gate (\ref{NOT}), an arbitrary basis
independent single qubit (unitary) gate for a real input qubit state
(\ref{states}) can be expressed as
\begin{equation}
 \label{single-gate}
U(\xi) \equiv \cos{\frac{\xi}{2}} \, I + \sin{\frac{\xi}{2}} \,
\textsl{NOT} \, ,
\end{equation}
where $I$ is the identity matrix and $\xi$ is a real free parameter
$0 \,\le\, \xi \,\le\, \pi$. This gate (\ref{single-gate}) performs
a rotation of the input qubit state (vector) on the angle $\xi$ in
the main circle. For example, a basis independent Hadamard gate
\cite{Pati:02}, that creates an equal superposition of a real qubit
and its orthogonal, corresponds to the rotation $U(\pi/2)$. In fact,
an arbitrary real qubit state can be obtained from a given real
state by means of a rotation (\ref{single-gate}) on some angle
$\xi$. From the symmetry of the Bloch sphere, moreover, it follows
that basis independent gates (\ref{NOT})-(\ref{single-gate}) can be
constructed for an arbitrary one-dimensional subspace of the Hilbert
qubit space.

Constructed single qubit gates (\ref{NOT})-(\ref{single-gate}) are
universal only for the real input qubit states (\ref{states}). It
means that these transformations are invariant with regard to the
basis rotation in the main circle only. Although the performance of
the gates (\ref{NOT})-(\ref{single-gate}) is restricted by the input
states from the set of real states, there are still infinitely many
states in the set. Moreover, for many computational tasks, the
application of the real states (\ref{states}) is indeed sufficient
\cite{Nielsen:00}.

In spite of the restriction discussed above, the gates
(\ref{NOT})-(\ref{single-gate}) may efficiently operate with input
mixed states. Suppose, a pseudo-pure state $\rho = \epsilon
\ket{\phi} \bra{\phi} + (1 - \epsilon) I$, where the pure state
$\ket{\phi}$, real state, is affected by the unitary gate
(\ref{single-gate}), i.e.
\begin{eqnarray}
 U \rho U^\dag  & = & U \left(\epsilon \ket{\phi} \bra{\phi} + (1 -
\epsilon) I \right) U^\dag
\nonumber\\[0.1cm]
& = & \epsilon \, U \ket{\phi} \bra{\phi} U^\dag + (1 - \epsilon) I
\, .
\end{eqnarray}
The unitary transformation does not change the purity parameter of
the input state, i.e. there is no loss of information, if the basis
independent gate (\ref{single-gate}) is applied to a pseudo-pure
state.

\subsection{\label{sec:3.3} Cloning-based architecture:
                            multi-qubit gates}

Having the single-qubit gates (\ref{NOT})-(\ref{single-gate}) let us
now construct a multi-qubit basis independent gate for real input
qubit states in order to complete the universal set of basis
independent gates. Let us start with the simplest multi-qubit gate
-- a two-qubit {\it Controlled-U} gate, assuming at the moment that
the input states of this gate are pure. Based on a straightforward
generalization of the classical Controlled-NOT gate, let us define
the quantum Controlled-U gate as follows. In a chosen computational
basis, the quantum Controlled-U gate leaves the state of the target
qubit unchanged, if the control qubit is given in the state
$\ket{0}$. If the control qubit is in the state $\ket{1}$, the gate
performs a unitary transformation $U$ on the target qubit. Since
after the transformation in the chosen basis, the output states of
the control and the target qubits are separable; for a basis
independent Controlled-U transformation, the output states must be
separable in any another basis. The state of the control qubit,
moreover, should not be changed after the transformation. Thereby,
for arbitrary superposed states of the control $\ket{\psi}_c$ and
the target $\ket{\chi}_t$ qubits, the basis independent Controlled-U
gate should perform some unitary transformation $\ket{f(\psi,
\chi)}_t$ on the target qubit leaving the control qubit without
changes, i.e.
\begin{eqnarray}
\label{U}
   \ket{\psi}_c \otimes \ket{\chi}_t & \longrightarrow &
   \ket{\psi}_c \otimes \ket{f(\psi, \chi)}_t \, .
\end{eqnarray}
The function $f(\psi, \chi)$ is related to the original state
$\ket{\chi}_t$ by a unitary transformation $\ket{f(\psi, \chi)}_t =
U(\psi)\:\ket{\chi}_t$ \cite{Siomau:10}.

An exact unitary transformation (\ref{U}) on arbitrary input qubit
states $\ket{\psi}_c$ and $\ket{\chi}_t$ is forbidden by the laws of
quantum mechanics. This was first pointed out by Pati \cite{Pati:02}
and is known today as {\it the general impossibility theorem}.
Indeed, to perform a transformation $U(\psi)$ on a state
$\ket{\chi}_t$, it is necessary to obtain some information about an
(unknown) input state $\ket{\psi}_c$ without changing the state.
This would be in conflict with the fundamental non-cloning principle
\cite{Wootters:82} which implies that no information can be obtained
from the state without changing it.

While the exact transformation (\ref{U}) on arbitrary input states
of the control and the target qubits does not exist, it is possible
to provide it approximately. A particular example of an approximate
transformation (\ref{U}) is the universal symmetric $1 \rightarrow
2$ quantum cloning machine (QCM) \cite{Buzek:96}. The QCM provides
approximate copying of an arbitrary input qubit state $\ket{\psi}_c$
to a prepared `blank' state $\ket{0}_t$, i.e.
\begin{eqnarray}
\label{QCM}
   \ket{\psi}_c \otimes \ket{0}_t \otimes \ket{A}_d & \longrightarrow &
   \ket{\psi}_c \otimes \ket{\psi}_t \otimes \ket{B}_d \, .
\end{eqnarray}
Here, we have introduced the initial $\ket{A}_d$ and the final
$\ket{B}_d$ states of an auxiliary system -- the cloning device
itself. The universal symmetric $1 \rightarrow 2$ QCM provides
transformation (\ref{QCM}) with an optimal fidelity $F = \spr{\psi}
{\rho |\, \psi} = 5/6$ between the input state $\ket{\psi}$ and each
copy $\rho$ \cite{Buzek:96}. This result can be generalized. For an
arbitrary input qubit state $\ket{\psi}_c$, the transformation
\begin{eqnarray}
\label{C-U-0}
   \ket{\psi}_c \otimes \ket{0}_t \otimes \ket{A}_d & \longrightarrow &
   \ket{\psi}_c \otimes U(\psi) \ket{0}_t \otimes \ket{B}_d \,
\end{eqnarray}
can be performed approximately with the optimal fidelity $F=5/6$
between each of the ideal outputs $\ket{\psi}_c$ and $U(\psi)
\ket{0}_t$ in right hand side and the corresponding actual results
of the transformation. Indeed, any transformation $U(\psi)
\ket{0}_t$ can be obtained as a sequence of copying $\ket{0}_t
\rightarrow \ket{\psi}_t$ (\ref{QCM}) and a unitary transformation
of the copy $U \ket{\psi}_t$. While the transformation of the copy
$U \ket{\psi}_t$ is not restricted by the laws of quantum mechanics,
the efficiency of the optimal transformation (\ref{C-U-0}) is
completely defined by the efficiency of the optimal cloning.
Moreover, there is a freedom in choice of the initial `blank' state
of the target qubit. If the input state of the target qubit is given
in the state $\ket{\chi}_t$, the transformation (\ref{C-U-0}) gives
rise to
\begin{eqnarray}
\label{C-U}
   \ket{\psi}_c \otimes \ket{\chi}_t \otimes \ket{A}_d & \longrightarrow &
   \ket{\psi}_c \otimes U(\psi) \ket{\chi}_t \otimes \ket{B}_d \, .
\end{eqnarray}
Therefore, any two-qubit Controlled-U transformation (\ref{C-U}) on
arbitrary input qubit states can be provided approximately with the
optimal fidelity $F = 5/6$ between the ideal outputs and the
corresponding actual outputs of the transformation.

So far we assumed that the input states of the universal
Controlled-U gate (\ref{C-U}) are arbitrary. If, however, the input
states of the control and the target qubits are not arbitrary but
taken from the the main circle of the Bloch sphere, the efficiency
of this transformation (\ref{C-U}) increases. It has been shown that
the optimal cloning (\ref{QCM}) of a real input qubit state can be
provided with fidelity $F = 1/2 + \sqrt{1/8} > 5/6$ between the the
copies and the input state \cite{Bruss:00}. Therefore, an
arbitrarily Controlled-U transformation (\ref{C-U}) can be provided
approximately on the input real states (\ref{states}) with the
optimal fidelity $F = 1/2 + \sqrt{1/8}$ between the ideal outputs
and the actual outputs of the transformation.

Recently, we showed an example of the universal Controlled-U gate
for real input qubit states -- the optimal basis independent C-NOT
gate \cite{Siomau:10}. Since the state of the input control qubit is
a superposition of the two basis states $\ket{0}_c$ and $\ket{1}_c$,
i.e.~$\ket{\psi} = \cos{\frac{\theta}{2}} \ket{0} \pm
\sin{\frac{\theta}{2}}\ket{1}$, the universal C-NOT gate for real
qubit states can be written explicitly for these basis states as
\begin{small}
\begin{eqnarray}
\label{CNOT-optimal-1}
   \ket{0}_c \ket{\chi_\pm}_t \ket{Q}_d
   &\longrightarrow & \left( \frac{1}{2} + \sqrt{\frac{1}{8}} \right)
   \ket{0}_c \ket{\chi_\pm}_t \ket{0}_d
\nonumber\\[0.1cm]
   & & \hspace*{-1.5cm} \:+\: \sqrt{\frac{1}{8}} \left( \ket{0}_c
   \ket{\chi_\pm^\bot}_t + \ket{1}_c \ket{\chi_\pm}_t \right) \,
   \ket{1}_d
\nonumber\\[0.1cm]
   & & \hspace*{-1.5cm} \:+\: \left( \frac{1}{2} - \sqrt{\frac{1}{8}}
   \right) \ket{1}_c \ket{\chi_\pm^\bot}_t \ket{0}_d \, ,
\\[0.2cm]
\label{CNOT-optimal-2}
   \ket{1}_c \ket{\chi_\pm}_t \ket{Q}_d
   &\longrightarrow & \left( \frac{1}{2} + \sqrt{\frac{1}{8}} \right)
   \ket{1}_c \ket{\chi_\pm^\bot}_t \ket{1}_d
\nonumber\\[0.1cm]
   & & \hspace*{-1.5cm} \:+\: \sqrt{\frac{1}{8}} \left( \ket{0}_c
   \ket{\chi_\pm^\bot}_t + \ket{1}_c \ket{\chi_\pm}_t \right) \,
   \ket{0}_d
\nonumber\\[0.1cm]
   & & \hspace*{-1.5cm} \:+\: \left( \frac{1}{2} - \sqrt{\frac{1}{8}}
   \right) \ket{0}_c \ket{\chi_\pm}_t \ket{1}_d \, ,
\end{eqnarray}
\end{small}
where $\ket{\chi_\pm}_t =  \cos{\frac{\phi}{2}} \ket{0}_t \pm
\sin{\frac{\phi}{2}} \ket{1}_t$ is the input state of the target
qubit. The state vector $\ket{Q}_d$ denotes the initial state of the
device that provides this transformation, while $\ket{0}_d$ and
$\ket{1}_d$ are the final states of the device. The output state
$\ket{\chi_\pm^\bot}_t$ of the target qubit is orthogonal to the
input target qubit state $\ket{\chi_\pm}_t$ and is obtained by
applying the NOT gate (\ref{NOT}) to the input state
$\ket{\chi_\pm}_t$, i.e. $\ket{\chi_\pm^\bot}_t = {\rm NOT}
\ket{\chi_\pm}_t$. Having a similar structure to the `equatorial'
QCM \cite{Bruss:00}, the gate
(\ref{CNOT-optimal-1})-(\ref{CNOT-optimal-2}) provides approximate
rotation of the target qubit $\ket{\chi_\pm}_t$ state vector on the
angle $\theta$ in the main circle, leaving the control qubit
untouched. The fidelity between the output and the ideal output of
the transformation equals $F = 1/2 + \sqrt{1/8}$. For more details
about the construction of the optimal basis independent C-NOT
transformation (\ref{CNOT-optimal-1})-(\ref{CNOT-optimal-2}) we
refer to \cite{Siomau:10}.

\begin{figure}
 \begin{center}
\includegraphics[scale=0.5]{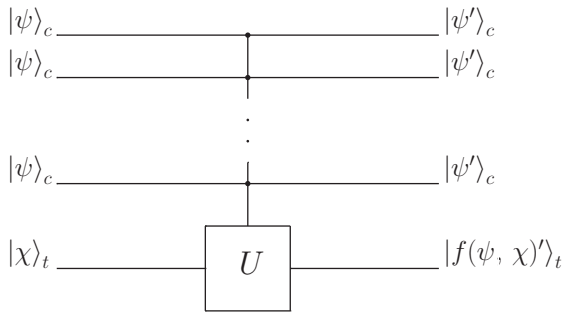}
\caption{The universal Toffoli gate (\ref{Toffoli}) with several
control qubits and one target qubit.} \label{fig:1}
 \end{center}
\end{figure}

Although the single qubit gates (\ref{NOT})-(\ref{single-gate}) with
the two-qubit C-NOT gate
(\ref{CNOT-optimal-1})-(\ref{CNOT-optimal-2}) form a universal set
of quantum gates that may support an arbitrary quantum computation
with real (pure) states of qubits, it is not possible to use these
gates for a realistic quantum computation. The reason for that is
the generally low fidelity $F = 1/2 + \sqrt{1/8}$ of the approximate
Controlled-U transformation (\ref{C-U}) for input real states. There
is, however, a native way to improve the fidelity of this
transformation. The approximate Controlled-U transformation
(\ref{C-U}) includes one control and one target qubit and is based
on the universal symmetric $1 \rightarrow 2$ cloning transformation.
In processing of classical algorithms there is a gate, called the
Toffoli gate, that includes some control qubits and one target qubit
\cite{Nielsen:00}. An extension of the optimal $1 \rightarrow 2$
cloning transformation to the case of the optimal $N \rightarrow M$
transformation is also well known \cite{Werner:98}. A combination of
the classical Toffoli gate with the universal symmetric $N
\rightarrow M$ QCM gives rise to a basis independent Toffoli gate,
which may be expressed as
\begin{eqnarray}
\label{Toffoli}
 \ket{\psi}_c^{\otimes N} \otimes \ket{\chi}_t
\otimes \ket{A}_d \longrightarrow \ket{\psi}_c^{\otimes N} \otimes
\ket{f(\psi, \chi)}_t \otimes \ket{B}_d \; .
\end{eqnarray}
This gate provides a specific transformation $\ket{f(\psi, \chi)}_t
= U(\psi)\:\ket{\chi}_t$ on a single target qubit in the presence of
$N$ control qubits and is shown schematically in Figure~\ref{fig:1}.

The basis independent transformation (\ref{Toffoli}) can be
performed approximately on arbitrary input qubit states with the
optimal fidelity (between each of the ideal outputs and the
corresponding actual outputs)
\begin{equation}
 \label{fidelity}
F_{N \rightarrow N+1}^u = 1 - \frac{1}{(N + 1) (N + 2)} \, ,
\end{equation}
which equals the optimal fidelity of the symmetric $N \rightarrow
N+1$ QCM \cite{Scarani:05, Werner:98}

\begin{figure}
 \begin{center}
 \includegraphics[scale=0.7]{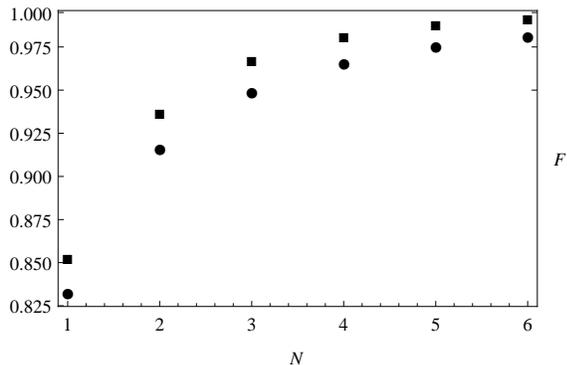}
\caption{Fidelity of the universal Toffoli gate (\ref{Toffoli}) as a
function of the number $N$ of control qubits. The circles show the
fidelity (\ref{fidelity}) of the gate for arbitrary input qubit
states. The squares give the upper bound for the fidelity
(\ref{fidelity pc}) for real input states.} \label{fig:2}
 \end{center}
\end{figure}

The fidelity (\ref{fidelity}) of the basis independent
transformation (\ref{Toffoli}) can be further improved taking into
account that the input states of this transformation are real qubit
states. It was shown that the optimal symmetric $N \rightarrow N+1$
cloning transformation for the real qubit states can be performed
with fidelity $F_{N \rightarrow N+1}^{pc}$, which is better than
$F_{N \rightarrow N+1}^u$ and has an upper bound given by
\begin{equation}
 \label{fidelity pc}
2 \frac{\sum_{i=0}^{N-1} \sqrt{C_i^N C_{i+1}^N}}{\sum_{j=0}^{N}
\sqrt{C_j^{N+1} C_{j+1}^{N+1}}} \geq F_{N \rightarrow N+1}^{pc} >
F_{N \rightarrow N+1}^u \, ,
\end{equation}
where $C_k^l$ denotes the binomial coefficient \cite{Bruss:00}. In
Figure~\ref{fig:2} the fidelity (\ref{fidelity}) of the approximate
basis independent Toffoli gate (\ref{Toffoli}) for arbitrary input
qubit states and the upper bound for the fidelity (\ref{fidelity
pc}) for input real states is displayed. The optimal fidelity $F_{N
\rightarrow N+1}^{pc}$ lies in between these values.


There is also another strategy to improve the fidelities of the
basis independent Controlled-U (\ref{C-U}) and Toffoli
(\ref{Toffoli}) gates. It is known that the optimal fidelity of a
cloning transformation increase for a small (in geometrical sense)
set of input states \cite{Fiurasek:03,Siomau:10a}. For example, one
may adopt the basis independent single qubit
(\ref{NOT})-(\ref{single-gate}) and the multi-qubit Toffoli
(\ref{Toffoli}) gates for input states from a small circle on the
Bloch sphere that is formed by a plane that crosses the sphere away
from its center. For input states from a small circle, a much higher
fidelity of the cloning transformation can be achieved in comparison
to input states from the main circle \cite{Fiurasek:03}.

Hitherto, in discussion of the optimal basis independent gates
(\ref{C-U}) and (\ref{Toffoli}), we hold the assumption that the
input states are real {\it pure} qubit states. This assumption can
be left now due to the work by Dang and Fan \cite{Dang:07}, who
derived optimal cloning transformations for initially mixed states
and showed that the optimal fidelities of these transformations
remain the same as in the case of initially pure states.

\subsection{\label{sec:3.4} Cloning-based architecture: some remarks}

The presented exact single-qubit gates
(\ref{NOT})-(\ref{single-gate}) with the approximate multi-qubit
Toffoli gate (\ref{Toffoli}) form a universal set of gates for real
states of qubits. Indeed, an arbitrary transformation of qubits
taken from the main circle of the Bloch sphere can be represented
through a sequence of the single- and multi-qubit basis independent
gates.

Although during the realization of the single Toffoli gate there is
the loss of information $ 1 - F $, it is always possible (by adding
control qubits and/or by making a proper choice of a set of input
qubits) to make this loss $ 1 - F $ less then a given value
$\delta$. In a particular algorithm, the overall fidelity between
the ideal output and the read-out of the algorithm can be estimated
as $F^\zeta$, where $F$ is the fidelity of a single N-qubit Toffoli
gate and $\zeta$ is an average number of the N-qubit Toffoli gates
acting on an input qubit.

During the construction of the basis independent transformations
(\ref{C-U}) and (\ref{Toffoli}) we have introduced an auxiliary
system in order to keep the discussion as general as possible.
However, the presence of an auxiliary system is not necessary for
some types of optimal basis independent transformations on real
qubit states \cite{Scarani:05,Niu:99}.

The present discussion covers the case of the deterministic quantum
computation within the circuit model. An attractive idea would be to
consider exact probabilistic quantum universal gates based on
probabilistic QCMs which allow one to perform \textit{exact} cloning
transformations with some probability \cite{Duan:98,Pati:99}.

It is also interesting to move beyond the circuit model and
consider, for example, optimal basis independent transformations
operating with a quantum system with a continuous spectrum
\cite{Scarani:05}.

There is also an important open question requiring an answer: ``What
is the role of entanglement in the implementation of the presented
basis independent gates?'' So far we are unable to answer this
question, since the role of entanglement is not clear in the cloning
process in general. In fact, the presence of entanglement in the
cloning process is widely confirmed \cite{Scarani:05,Bruss:03}. It
is also known, however, that no entanglement is required for optimal
cloning in the limit of large dimensions \cite{Scarani:05}.

\section{\label{sec:4} Physical realization of a quantum computer}

Inasmuch as we hope that a quantum computer will not be just an
amusing occurrence in quantum theory but will actually be realized
as a practical powerful device, we need to think about the simple
and manageable {\it physical realization} of a quantum computer.
Since an absolute majority of experiments have been devoted to a
physical realization of a SQC, let us first briefly overview the
most important (from our point of view) recent experimental
achievements in the realization of SQC.

Several experimental realizations of the basic element for the
scheme for linear optics quantum computation \cite{Knill:01}, the
C-NOT gate (\ref{standart-CNOT}), have been already reported
\cite{Kok:07}. However, the gate that follows the original setup by
Knill, Laflamme and Milburn has been just recently demonstrated to
have an average fidelity $\overline{F}= 0.82 \pm 0.01$ between the
output and the ideal output, which is indeed far from the
theoretically predicted unit fidelity \cite{Okamoto:10}.

Significant progress has been achieved in the realization of basic
gates for a SQC with trapped ions \cite{Haeffner:08}. To our
knowledge the best realization of a C-NOT gate has been reported to
have an average fidelity $\overline{F}= 0.940 \pm 0.004$ between the
output and the ideal output \cite{Home:09}. However, the fidelity is
still quite far from the unit fidelity.

We also would like to mention the recent experimental achievements
in the realization of a scheme for a SQC that is beyond the circuit
model, the one-way quantum computing \cite{Raussendorf:01}, in order
to envelop different physical realizations of a SQC. In this scheme
the algorithm realization is achieved by a sequence of single qubit
measurements on an initially multi-qubit entangled state, a cluster
state. In spite of a very efficient realization of this scheme on
three qubits \cite{Hamel:09}, the production of multi-qubit cluster
states remains a serious practical problem. After almost ten years
of the presentation of the scheme for the one-way quantum computing,
only a six-qubit cluster state is available in a lab \cite{Lu:07}.

In spite of profound progress in experimental realization of SQC
during recent years, there are still many practical problems in each
of mentioned realizations of a SQC \cite{Ladd:10,Home:09}. On this
background the realization of the cloning-based scheme for quantum
computing looks quite promising. An optical implementation of the
universal $1 \rightarrow 2$ QCM for an arbitrary input qubit state
\cite{Buzek:96} based on parametric down-conversion has been
demonstrated to have fidelity $0.810 \pm 0.008$ \cite{DeMartini:02}
which is in good agreement with the theoretical prediction $5/6 =
0.833$. Recently this result has been further improved
\cite{Pittman:05}. Another physical realization of the universal QCM
was presented in \cite{Fasel:02}; using optical fibers doped with
erbium ions, this transformation was shown to have fidelity $F
\approx 0.82$ which is again in good agreement with the theoretical
prediction. Also, several realistic theoretical schemes for the
physical realization of a QCM on atoms in a cavity have been
recently proposed \cite{Milman:03,Yang:08,Fang:09}. However, to our
knowledge no experimental results are available.

It is, however, worth noticing that most experiments on the
realization of a QCM have been devoted to universal symmetric $1
\rightarrow N$ QCMs where $ N = 2,3...$ \cite{Scarani:05}. No
attention has been paid to realization of universal symmetric $N
\rightarrow N+1$ QCMs. Therefore, it is hard to judge whether such a
cloning machine and corresponding optimal basis independent Toffoli
gate (\ref{Toffoli}) can be efficiently realized in practice.

\section{\label{sec:5} Summary}

After a brief discussion of the efficiency of some quantum
algorithms \cite{Knill:98,Shor:94,Grover:97,Deutsch:92,Simon:97},
assuming that the input states are mixed, we have presented a new
architecture for a quantum computer. This architecture includes a
universal set of basis independent singe-qubit
(\ref{NOT})-(\ref{single-gate}) and multi-qubit Toffoli
(\ref{Toffoli}) gates that need to be applied to the input qubit
states taken from the main circle on the Bloch sphere. Although the
basis independent Toffoli gate can be applied on input qubits only
approximately, the fidelity of the approximate transformation can be
done as close to unity as required by adding control qubits and/or
by making a proper choice of a set of input qubits. Since the
suggested basis independent Toffoli gate is based on a universal
quantum cloning machine, we briefly reviewed, in addition, recent
experimental achievements in the realization of quantum cloning
machines. Finally, we would like to note that this paper presents
just a particular way of thinking about quantum computing with mixed
states that may be useful in the development of new algorithms and
in future experiments.

\begin{acknowledgments}
 This work was supported by the Deutsche Forschungsgemeinschaft.
\end{acknowledgments}

\end{document}